\documentclass{amsart} 
\usepackage{amsmath,euscript,graphicx,xspace,natbib}
\usepackage{nicefrac}
\usepackage{rotating, wasysym} 

\newcommand{\agent}              {\ensuremath{{a}}\xspace}

\newcommand{\step}               {\ensuremath{{t}}\xspace}
\newcommand{\sector}            {\ensuremath{i}\xspace}
\newcommand{\sectors}           {\ensuremath{n}\xspace}

\newcommand{\as}                 {_\agent(\step)}

\newcommand{\is}                  {_\sector(\step)}
\newcommand{\ias}                {_{\sector\agent}(\step)}
\newcommand{\iar}                {_{\sector\agent}(\step-1)}
\newcommand{\ar}                  {_\agent(\step-1)}

\newcommand{\strategy}        {\ensuremath{{s}}\xspace}
\newcommand{\income}         {\ensuremath{{Y}}\xspace}
\newcommand{\price}              {\ensuremath{p}\xspace}
\newcommand{\growth}         {\ensuremath{{\gamma}}\xspace}
\newcommand{\coefficient}     {\ensuremath{{\alpha}}\xspace}
\newcommand{\capital}           {\ensuremath{{K}}\xspace}
\newcommand{\scaling}         {\ensuremath{{\beta}}\xspace}
\newcommand{\deprecation}   {\ensuremath{{\delta}}\xspace}

\newcommand{\limit}                {\ensuremath{\lim_{\step\to\infty}}\xspace}

\newcommand{\response}         
{\ensuremath{\prod_\sector\strategy\ias^{\coefficient_\sector}}\xspace}

\begin{document}

\raggedbottom 

\title[Convergence of Income Growth Rates]{Convergence of Income Growth Rates in\\ Evolutionary Agent-Based Economics}
\date{}
\author{Volker Nannen}

\maketitle


\vspace*{-6mm}
\begin{center}
\sc\small
Departament de Ci\`encies Matem\`atiques i Inform\`atica\\[1mm]
Universitat de les Illes Balears
\end{center}
\medskip

\begin{abstract}
We consider a heterogeneous agent-based economic model where economic agents 
have strictly bounded rationality and where income allocation  
strategies evolve through selective imitation. Income is calculated by a 
Cobb-Douglas type production function, and selection of strategies for
imitation depends on the income growth rate they generate. We show that
under these conditions, when an agent adopts a new strategy, the effect on its
income growth rate is immediately visible to other agents, which allows a
group of imitating agents to quickly adapt their strategies when needed. 
\end{abstract}

\section{Introduction}
In biology, fitness describes the capability of an individual of a certain
genotype to reproduce. In analogy, fitness in this context of evolutionary
economics describes the likelihood that an agent with a certain behavior is
imitated by other agents. If the likelihood that an agent with a certain
investment strategy is imitated depends on the relative income growth rate of
the agent, the functional relationship between investment strategies and
income growth rates is central to the understanding of the emerging
evolutionary dynamics, and can be used to develop new economic policy tools 
\citep[see for example][]{Nannen-vandenBergh:2010}.
Here we study how the income growth rate of an agent stabilizes after a change
of strategy, i.e., how it converges to the equilibrium growth rate that an
imitating agent realizes if it holds on to a particular investment strategy
and when prices are stable. In particular, we want to know if there is
significant delay between the adoption of a new strategy and the expression
of this strategy in terms of income growth.

\section{Strategies, growth and production}
In this model, agents represent firms. Each agent formulates an investment
strategy that specifies how current income is invested in a finite number of
capital sectors, e.g., security, machinery, or sales infrastructure. The
returns for each agent are then calculated from standard economic growth and
production functions. We do not model savings, and so the income growth rate
is also the growth rate of returns on investments. Some allocations give
higher returns than others, and the goal of the agents is to find a strategy
that can realize a high level of individual welfare. Agents have bounded
rationality and limited information. They observe the investment strategy and
resulting income growth rate of other agents, and imitate a successful
investment strategy, possibly with some variation. This constitutes a
collective learning process, where strategies evolve by selection and
variation.

Let \sectors be the number of available investment sectors. Formally, the
investment strategy $\strategy\as$ of agent \agent at time \step can be
defined as an \sectors-dimensional vector 
\begin{equation}\label{eq:strategy}
  \strategy\as = \left[0,1\right]^\sectors,\quad \sum_\sector\strategy\ias=1.
\end{equation}
Let $\income\as$ be the income of agent \agent at time \step. The partial
strategy $\strategy\ias$---which is the $i^{th}$ element of a
strategy---determines the fraction $\strategy\ias\income\ar$ of income that
agent \agent invests in sector \sector at time \step. As agents allocate all
of their income over the available sectors, the partial strategies must be
non-negative and sum to one. The set of all possible investment strategies
forms an $\sectors-1$ dimensional simplex that is embedded in
\sectors-dimensional Euclidean space.

We use standard economic growth and production functions to describe how
capital accumulates in each sector and contributes to income. These functions
are not aggregated: growth and returns are calculated independently for each
agent. We consider invested capital to be non-malleable: once invested it
cannot be transferred between sectors. Capital accumulation in each sector
depends on the sector specific investment of each agent, a dynamic price
$\price\is$, and the global deprecation rate $\deprecation$. Deprecation is
assumed to be equal for all sectors and all agents. The difference equation
for non-aggregate growth per sector is 
\begin{equation}\label{eq:sector}
  \capital\ias = \frac{\strategy\ias}{\price\is}\income\ar +
 (1-\deprecation)\capital\iar.
\end{equation}

To calculate the income $\income\as$ from the capital that agent \agent has
accumulated per sector, we use an $n$-factor Cobb-Douglas production function
with constant elasticity of substitution,
\begin{equation}\label{eq:production}
 \income\as = \scaling \prod_\sector{\capital\ias}^{\coefficient_\sector},
\end{equation}
where \scaling is a scaling factor that limits the maximum possible income
growth rate. The relative contribution of each sector to production is
expressed by a vector of non-negative production coefficients
$\coefficient = \langle \coefficient_1 \ldots \coefficient_\sectors \rangle$.
To enforce constant returns to scale, all production coefficients are
constraint to add up to one,
\begin{equation}\label{eq:coefficient}
  \coefficient = \left[0,1\right]^\sectors,
  \quad \sum_\sector \coefficient\is=1.
\end{equation}
Similar to the strategy space, the set of all possible vectors of production
coefficients is an $\sectors-1$ dimensional simplex that is embedded in
\sectors-dimensional Euclidean space. 

\section{The equilibrium growth rate} To find the equilibrium growth rate, we
start with an analysis of the equilibrium ratio of sector specific capital to
income that will be achieved if an agent holds on to a particular strategy.
The difference equation of this ratio is

\begin{equation}\label{eq:ratio}
 \begin{aligned}
  \frac{\capital\ias}{\income\as}
   =\;&\frac{\nicefrac{\strategy\ias}{\price\is}\income\ar
       \;+\;(1-\deprecation)\capital\iar}{(\growth\as+1)\;\income\ar}\\[3mm]
   =\;&\frac{\strategy\ias/\price\is}{\growth\as+1} +
       \frac{1-\deprecation}{\growth\as+1}\;
       \frac{\capital\iar}{\income\ar}.
 \end{aligned}
\end{equation}

\medskip\noindent This dynamic equation is of the form $$x(\step) =
a+bx(\step-1),$$ which under the condition $0\!\le\!b\!<\!1$ converges
monotonously to its unique stable equilibrium at $$\limit x(\step) \!=\!
a/(1-b).$$ This condition is fulfilled here: investment is always
non-negative and sector specific capital cannot decrease faster than
$\deprecation$. With constant returns to scale, income cannot decline faster
than capital deprecation, i.e., $\growth_\agent \!\ge
\!-\deprecation$. For the moment, let us exclude the special case
$\growth_\agent \!=\! -\deprecation$. Then, considering that $0 \!<\!
\deprecation \!\le\! 1$, we have the required constraint
\begin{equation}\label{eq:equilibrium-condition}
  0\le\frac{1-\deprecation}{\growth\as+1}<1.
\end{equation}
We conclude that the ratio of capital to income converges to 
\begin{equation}\label{eq:equilibrium-ratio}
 \begin{aligned}
  \limit\frac{\capital\ias}{\income\as} 
  =& \limit\frac{\strategy\ias/\price\is}{\growth\as+1}/\left(1-
     \frac{1-\deprecation}{\growth\as+1}\right)\\[2mm]
  =& \limit\,\frac{\strategy\ias/\price\is}{\growth\as+\deprecation}.
 \end{aligned}
\end{equation}
The existence of this limit depends on the behavior of $\price\is$. If prices
converge, equation \ref{eq:equilibrium-ratio} describes a unique stable
equilibrium to which the ratio of capital to income converges monotonously. We
ignore the limit notation and combine equation \ref{eq:equilibrium-ratio} with
equation \ref{eq:production} to calculate income at equilibrium as
\begin{equation}\label{eq:equilibrium-income}
 \begin{aligned}
  \income\as &= \scaling\, 
  \prod_\sector\left(\frac{\strategy\ias/\price\is}
  {\growth\as+\deprecation}\income\as\right)^{\coefficient_\sector}\\[2mm]
  &= \frac{\scaling\,\income\as}{\growth\as+\deprecation}\,
  \prod_\sector \price\is^{-\coefficient_\sector}\,
  \prod_\sector \strategy\ias^{\coefficient_\sector}.
 \end{aligned}
\end{equation}
We can now solve for $\growth\as$ to derive the equilibrium growth rate
\begin{equation}\label{eq:equilibrium-growth}
  \growth\as=\scaling\, \prod_\sector \price\is^{-\coefficient_\sector}\, 
  \response-\deprecation.
\end{equation}

Let us return to the special case $\growth_\agent =-\deprecation$.
According to equation \ref{eq:sector}, capital per sector decreases at the
deprecation rate \deprecation only when it receives zero investment, and it
cannot decrease faster. With constant elasticity of substitution, a growth of
$\growth_\agent =-\deprecation$ is only possible if every sector with a
positive production coefficient receives zero investment. This implies
$\strategy\ias=0$ for at least one partial strategy, and so equation
\ref{eq:equilibrium-growth} holds also for the special case
$\growth_\agent=-\deprecation$.

\section{Relative order of strategies}
The scaling factor \scaling, the price \price and the deprecation rate
\deprecation are monotonous transformations of the equilibrium growth rate in
equation \ref{eq:equilibrium-growth}. They do not affect the order of
strategies with respect to the income growth rate at equilibrium.
Whether one investment strategy results in a higher income growth rate at
equilibrium than another investment strategy depends solely on the term
\response. Further to this, a set of investment strategies with identical
equilibrium growth rate, say $\growth'$, forms a contour hyper surface in the
strategy simplex. All strategies that are enveloped by this hyper surface have
a higher equilibrium growth rate $\growth\as\ge\growth'$. This inner set is
convex \citep[for a related proof see][]{Beer:1980} and satisfies 
\begin{equation}
      \prod_\sector\strategy\ias^{\coefficient_\sector}
  \ge \frac{\growth'+\deprecation}{\scaling} \prod_\sector
\price\is^{\coefficient_\sector}.
\end{equation} 

Maximizing this type of equilibrium growth rate poses no challenge to a 
(collective) learning mechanism. It has a single global maximum at $\strategy\as \!=\!
\coefficient$, no local maxima, and a distinct slope that increases away
from the maximum, allowing even the simplest of hill climbing algorithms to
find and approach it. Learning mechanisms will differ mostly in the speed of
convergence and the degree of fine tuning at the optimum. 

\section{Convergence}
Numerical simulations reveal that after an agent has changed its investment 
strategy, its income growth rate approaches the equilibrium growth rate of the
new strategy always monotonously from above. This might be obvious when an agent 
exchanges a superior strategy for an inferior strategy. But it is also true when 
an agent exchanges an inferior for a superior strategy, because the change in the
distribution of an agent's total capital over the sectors is largest
immediately after a change in the allocation of investments, with a strong
effect on the income growth rate. Even when an agent switches between
two strategies that have the same equilibrium growth rate, the agent will
temporarily increase its income growth rate and approach the equilibrium
growth rate from above. That is, the agent will sustain an equilibrium growth rate
that is higher than (or equal to) the equilibrium growth rate of each of the two
strategies on their own. This is due to the convex shape of the equilibrium
growth
rate. Any convex combination of two strategies with equal equilibrium
growth rate must have an equilibrium growth rate at least as high as that 
of the two combining strategies. The temporal effect of a switch between two
strategies is similar to that of a convex combination of the two strategies.

This peculiar convergence behavior introduces an intrinsic bias into
the evolutionary process that favors an agent that has just changed its
strategy. Figure \ref{figure:convergence-behavior} visualizes the behavior of
the income growth rate of a single agent that changes its strategy several
time over the course of 500 time steps. In this example the economy is
calibrated such that the equilibrium growth rate of the optimal strategy is
1.85 percent per year. Each random strategy is another imitation (i.e., copy)
of the optimum strategy with an imitation error that follows a Gaussian
distribution ${\mathcal N}(0,0.02)$. Immediately after the imitation, the
income growth rate of the imitating agent always exceeds the equilibrium
growth of its new (and nearly optimal) strategy by a significant margin.
Occasionally the actual growth exceeds even the equilibrium growth rate of
the optimal strategy, which means that if at that moment another agent has to
choose whether to imitate this agent or an agent that is using the optimal
strategy, it is likely to imitate this agent. 

\newcommand{\myfigure}[4]{\small
  \parbox{\textwidth}{\centering\hspace{6mm} #3\\[0mm]\sffamily
  \begin{sideways}\ \ \parbox{.29\textwidth}{\centering #1}\end{sideways}
  \includegraphics[width=.95\textwidth,height=.34\textwidth]
  {#4.eps}\\\hspace{6mm} #2\\[2mm]}}
\begin{figure}[t]
\begin{center}
  \myfigure{income growth rate}{time steps}{Comparing the real income growth
rate to the equilibrium growth rate}{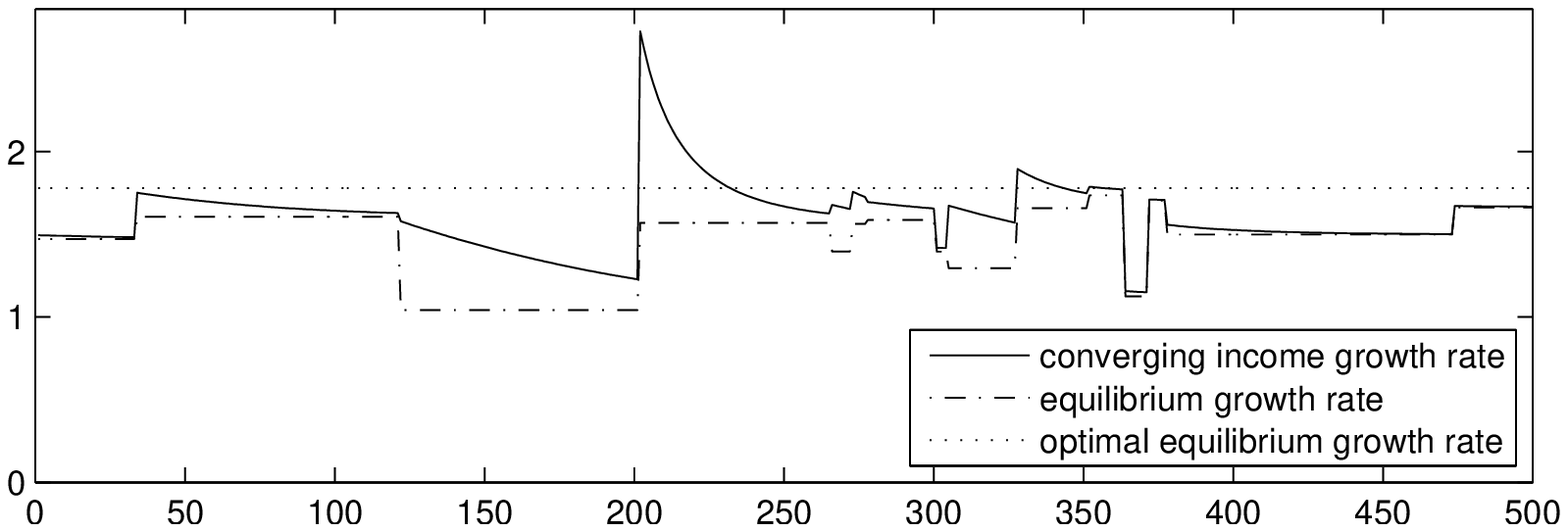}\\[3mm]
  \myfigure{excess growth rate}{time steps}{converging growth rate
minus equilibrium growth rate}{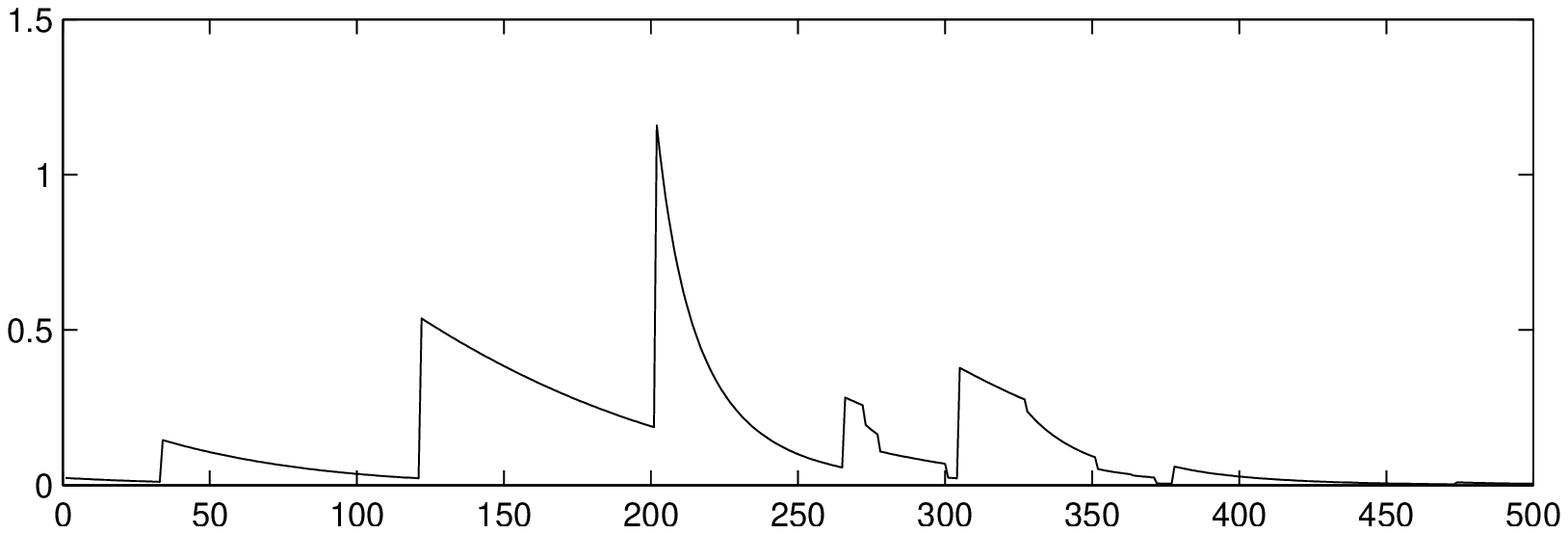}
\caption{Income growth rate of a single agent during 500 time steps. The agent
changes its strategy several times.}
\label{figure:convergence-behavior}
\end{center}
\vspace*{6mm}
\end{figure}

\newpage

\section{Conclusion}
We conclude that if imitating agents base the decision from which agent to
imitate on the income growth of their fellow agents, there is no delay between
the adoption of a strategy (the genotype) and expression of this strategy in
terms of economic performance (the phenotype), which is of great advantage to
an evolutionary process that is based on imitation. However, the convergence
behavior introduces significant biased in favor of those agents that have
only recently changed their strategy. This might prevent an agent population
to fully converge on the optimal strategy, and preserve variation in the pool
of strategies.

\bibliographystyle{elsart-harv}

\end{document}